\documentclass[aps,prd,nofootinbib,twocolumn,superscriptaddress,preprintnumbers,balancelastpage,longbibliography]{revtex4-2}
\usepackage{aas_macros}
\usepackage[utf8]{inputenc}
\usepackage{amsmath,amssymb,mathtools,bm}
\usepackage[section]{placeins}
\usepackage{graphicx, color, hepunits}
\usepackage[dvipsnames]{xcolor}
\usepackage{float}
\usepackage{multirow}
 \usepackage{hyperref} 
\hypersetup{
    colorlinks=true,       
    linkcolor=blue,        
    citecolor=blue,        
    filecolor=magenta,     
    urlcolor=blue          
}
\usepackage[utf8]{inputenc}
\usepackage[english]{babel}
\usepackage{lipsum}

\usepackage{tensor}
\usepackage{slashed}
\usepackage{lineno}

\newcommand{\gagg}{{g_{a\gamma\gamma}}}

\newcommand{\gaee}{{g_{aee}}}

\newcommand{\gagggaee}{g_{a \gamma \gamma}\times g_{aee}}

\newcommand{\farcs}{\mbox{\ensuremath{.\!\!^{\prime\prime}}}}

\begin{document}

\title{Search for Axions from Magnetic White Dwarfs with Chandra}

\author{Orion Ning}
\email{orion.ning@berkeley.edu}
\affiliation{Berkeley Center for Theoretical Physics, University of California, Berkeley, CA 94720, U.S.A.}
\affiliation{Theoretical Physics Group, Lawrence Berkeley National Laboratory, Berkeley, CA 94720, U.S.A.}

\author{Christopher Dessert}
\email{cdessert@flatironinstitute.org}
\affiliation{Center for Computational Astrophysics, Flatiron Institute, New York, NY 10010, USA}

\author{Vi Hong}
\affiliation{Berkeley Center for Theoretical Physics, University of California, Berkeley, CA 94720, U.S.A.}
\affiliation{Theoretical Physics Group, Lawrence Berkeley National Laboratory, Berkeley, CA 94720, U.S.A.}

\author{Benjamin R. Safdi}
\email{brsafdi@berkeley.edu}
\affiliation{Berkeley Center for Theoretical Physics, University of California, Berkeley, CA 94720, U.S.A.}
\affiliation{Theoretical Physics Group, Lawrence Berkeley National Laboratory, Berkeley, CA 94720, U.S.A.}

\date{\today}

\begin{abstract}
Low mass axion-like particles could be produced in abundance within the cores of hot, compact magnetic white dwarf (MWD) stars from electron bremsstrahlung and converted to detectable X-rays in the strong magnetic fields surrounding these systems. 
In this work, we constrain the existence of such axions from two dedicated {\it Chandra} X-ray observations of $\sim$40 ks each in the energy range $\sim$1 -- 10 keV towards the magnetic white dwarfs (MWDs) WD 1859+148 and PG 0945+246. 
We find no evidence for axions, which constrains the axion-electron times axion-photon coupling
to $|g_{a\gamma \gamma} g_{aee}| \lesssim 1.54 \times 10^{-25}$ ($3.54 \times 10^{-25}$) GeV$^{-1}$ for PG 0945+246 (WD 1859+148) at 95\% confidence for axion masses $m_a \lesssim 10^{-6}$ eV. We find an excess of low-energy X-rays between 1 -- 3 keV for WD 1859+148 but determine that the spectral morphology is too soft to arise from axions; instead, the soft X-rays may arise from non-thermal emission in the MWD magnetosphere.
\end{abstract}
\maketitle

\section{Introduction}

Axion-like particles are beyond the Standard Model candidates that are motivated by physics at high energy scales such as string theory compactifications~\cite{Green:1984sg,Witten:1984dg,Svrcek:2006yi,Arvanitaki:2009fg}.
Axions interact with the Standard Model through dimension-five and higher operators suppressed by the high scale $f_a$ of the ultraviolet (UV) completion, leading to faint but potentially observable signatures in laboratory and astrophysical probes (see~\cite{Hook:2018dlk,DiLuzio:2020wdo,Safdi:2022xkm,Adams:2022pbo,OHare:2024nmr} for reviews). Axions that couple to quantum chromodynamics (QCD) acquire mass contributions from non-perturbative QCD effects; these axions may solve the strong-{\it CP} problem~\cite{Peccei:1977ur,Peccei:1977hh,Weinberg:1977ma,Wilczek:1977pj}. Axion-like particles do not couple to QCD but may interact with the rest of the Standard Model. Their masses need not be correlated with  their coupling constants and could be ultra-light.  In this work we constrain light axion-like particles (hereafter referred to as axions) by using dedicated observations of magnetic white dwarfs (MWDs) in the X-ray band.
\begin{figure}[!htb]
\centering
\includegraphics[width=0.49\textwidth]{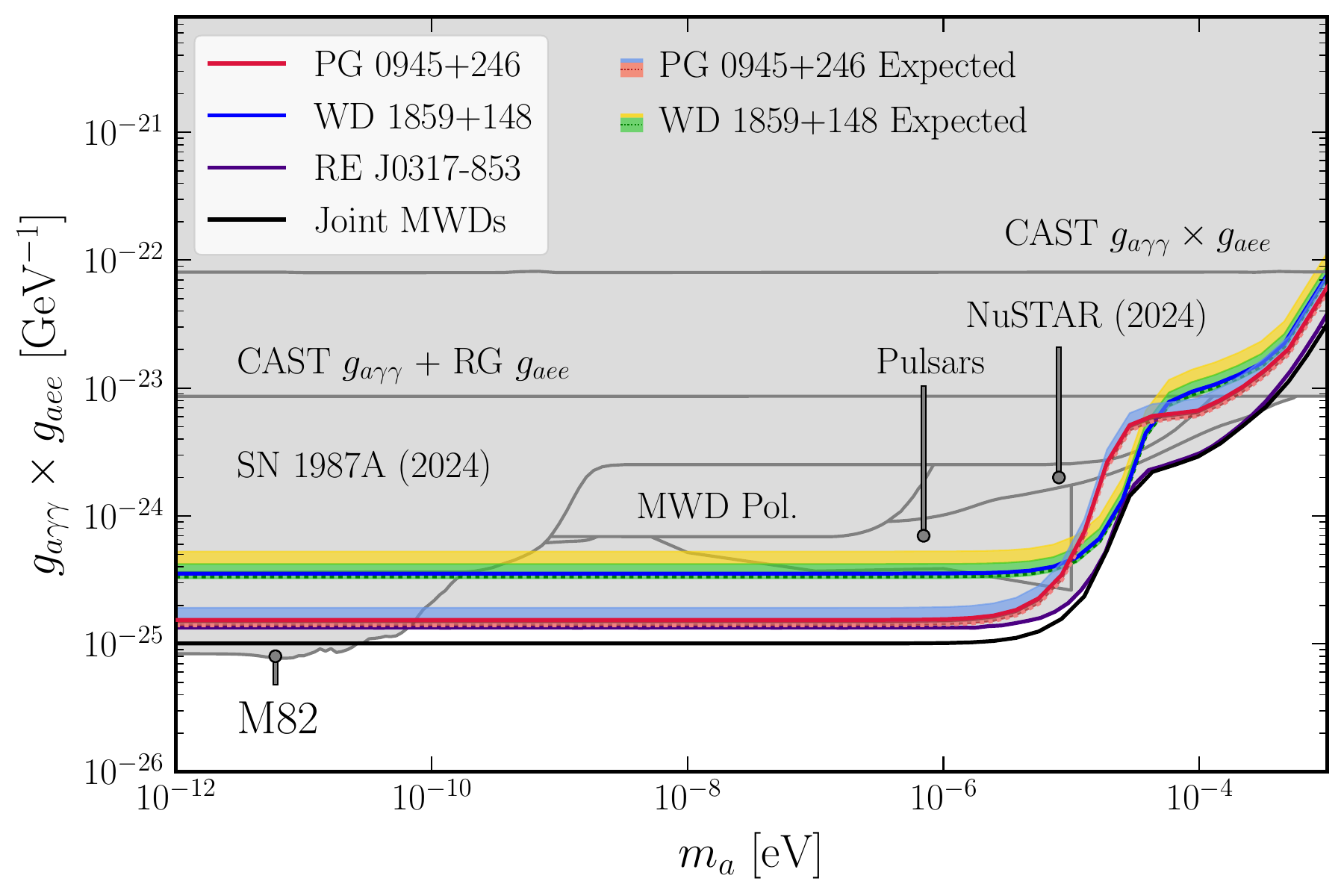}
\vspace{-0.4cm}
\caption{ The nonobservation of axion-induced X-ray signals in the MWDs PG 0945+246 and WD 1859+148 constrain $|\gagggaee| \lesssim 1.54 \times 10^{-25}$ GeV$^{-1}$ and $|\gagggaee| \lesssim 3.54 \times 10^{-25}$ GeV$^{-1}$, respectively, at 95\% confidence at low $m_a$. For both MWDs, we also illustrate the $1/2\sigma$ containment intervals for the expected 95\% one-sided upper limits under the null hypothesis. We show previous constraints in gray (see main text)~\cite{2013ApJS..208....4P, Noordhuis:2022ljw, Ruz:2024gkl, Ning:2024eky,Manzari:2024jns, Dessert:2022yqq, Anastassopoulos:2017ftl, Capozzi:2020cbu}, and in particular we highlight the limit obtained from a similar search to that performed here using {\it Chandra} data towards the MWD RE J0317-853~\cite{Dessert:2021bkv}, which can be combined with the upper limits from the MWDs in this work to form a joint limit, also shown.
}
\label{fig:gaeegagg}
\end{figure}

It is well-known that axions could be produced within the hot cores of MWDs from electron bremsstrahlung and converted to X-rays in the surrounding magnetic fields~\cite{Dessert:2019sgw,Dessert:2021bkv}. 
This search uses the following terms in the axion ($a$) effective field theory
\begin{equation}
    \mathcal{L} \supset -\frac{1}{4} \gagg a F_{\mu \nu} \Tilde{F}^{\mu \nu} + \frac{\gaee}{2 m_e} (\partial_{\mu}a) \Bar{e} \gamma^{\mu} \gamma_5 e \,,
\end{equation}
where $\gagg$ is the axion-photon coupling, $\gaee$ is the axion electron coupling, $F_{\mu \nu}$ is the quantum electrodynamics (QED) field strength, $e$ is the electron field, and $m_e$ is the electron mass.  Axions could be produced through interactions involving $g_{aee}$ with energies of order the MWD core temperature $T \sim {\rm keV}$ and efficiently converted to X-rays using $g_{a\gamma\gamma}$ in the MWD magnetosphere.
A dedicated search was performed using 40 ks of {\it Chandra} data 
towards the MWD target RE J0317-853~\cite{Dessert:2021bkv}; no evidence for X-ray emission was found, leading to the current leading upper limit illustrated in Fig.~\ref{fig:gaeegagg}.  That figure illustrates the upper limits on the coupling constant combination $|g_{a\gamma\gamma} \times g_{aee}|$ as a function of the axion mass $m_a$. 
We include the direct constraint on $|g_{a\gamma\gamma} \times g_{aee}|$ from CAST~\cite{2013JCAP...05..010B} and additionally show the upper limits derived from combining the leading upper limits on $g_{a\gamma\gamma}$ alone from a variety of astrophysical probes~\cite{Noordhuis:2022ljw, Ruz:2024gkl, Ning:2024eky,Manzari:2024jns, Dessert:2022yqq, Anastassopoulos:2017ftl} with the leading upper limits on $g_{aee}$ from stellar cooling~\cite{Capozzi:2020cbu}.

In this work we collect and analyze \textit{Chandra} data towards two new MWDs: WD 1859+148 and PG 0945+246. A priori, these are promising targets to probe axion emission due to their relatively high predicted core temperatures and magnetic fields, which are both likely to exceed those of RE J0317-853~\cite{Caiazzo:2021xkk, Ferrario:2015oda, Camisassa:2022pet}.
As we show in this article, we find no evidence for axions from these new observations and end up with comparable sensitivity relative to the RE J0317-853 analysis.
There are a few reasons why our analyses do not give substantially improved sensitivity relative to the analysis of RE J0317-853 {\it Chandra} data. One of the MWDs shows an excess of low-energy X-rays (between 1 - 3 keV). As we discuss, this excess appears too soft to be explained by axions. (Ref.~\cite{Bamba:2024nwd} analyzed these data and suggested this excess may arise from non-thermal activity in the MWD magnetosphere.) We thus exclude the 1 - 3 keV energy range from our analysis of data from that MWD, which lowers our sensitivity. Secondly, we note that the {\it Chandra} effective area has decreased markedly since the 2020-12-18 observation of RE J0317-853, which results in fewer expected signal counts for the same signal flux today relative to 2020.  Still, given the similarity of our results to those from RE J0317-853 and the inherent astrophysical uncertainties due to modeling the WD interior that affect these analyses, our work adds strength to the overarching result that axions with $m_a \lesssim 10^{-6}$ eV and $|\gagg \times \gaee| \gtrsim 2 \times 10^{-25}$ GeV$^{-1}$ are strongly excluded.  

\section{Data collection and analysis} 

\begin{figure}[!htb]
\centering
\includegraphics[width=\columnwidth]{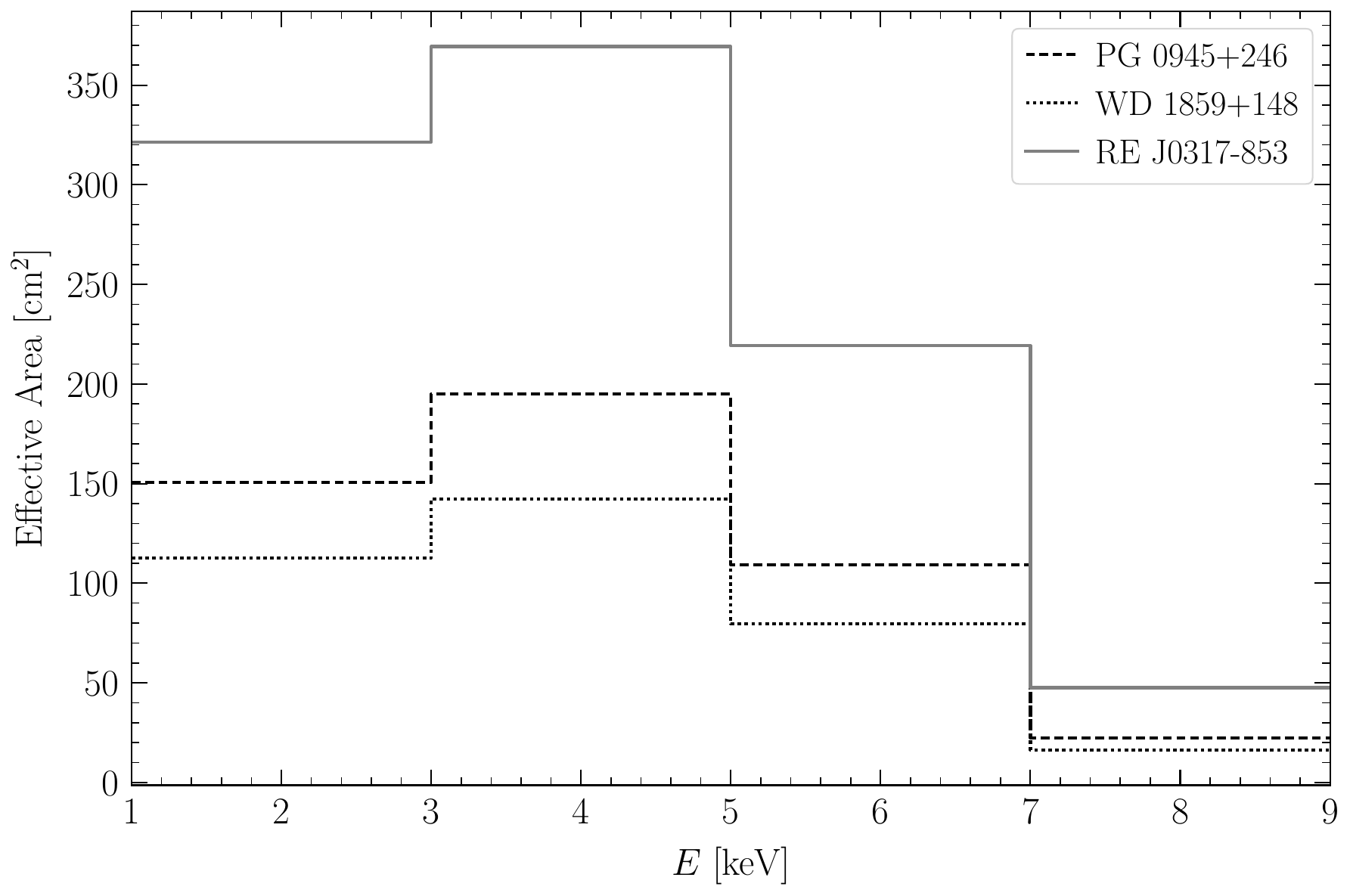}
\caption{ A comparison of the \textit{Chandra} effective area over our energy analysis range, across the observational epochs corresponding to PG 0945+246, WD 1859+148, and RE J0317-853~\cite{Dessert:2021bkv}. As discussed in the main text, the degradation of the \textit{Chandra} instrument over the past few years impacts the effective area, which is clearly apparent here. }
\label{fig:eff_areas}
\end{figure}

We observed the MWD WD 1859+148 on 2022-12-18 using the {\it Chandra} ACIS-I instrument with no grating for a total of 39.31 ks (PI Safdi, observation IDs \texttt{26496}, \texttt{27596}, and \texttt{27597}), and the MWD PG 0945+246 on 2023-12-16 for 36.61 ks (observation IDs \texttt{26497}). After data reduction and stacking -- see App.~\ref{app:data} -- we produce pixelated counts and exposure maps in four energy bins from 1 to 9 keV of width 2 keV each and into pixels of size $0\farcs492$.  Note that the pixel sizes are near the angular resolution of the instrument; the 68\% containment radius of the point spread function (PSF) is around $0\farcs5$ over the energy range of interest. The build-up of a contaminant has reduced the effective area of the instrument below 2 keV to near-zero. 
In Fig.~\ref{fig:eff_areas} we compare the effective area between the observations of RE J0317-853, WD 1859+148, and PG 0945+246.  The 2020 observation of RE J0317-853 was thus seen to take place with the instrument in a superior state relative to the 2022 and 2023 observations analyzed in this work. 

\begin{figure*}[!htb]
\centering
\includegraphics[width=0.49\textwidth]{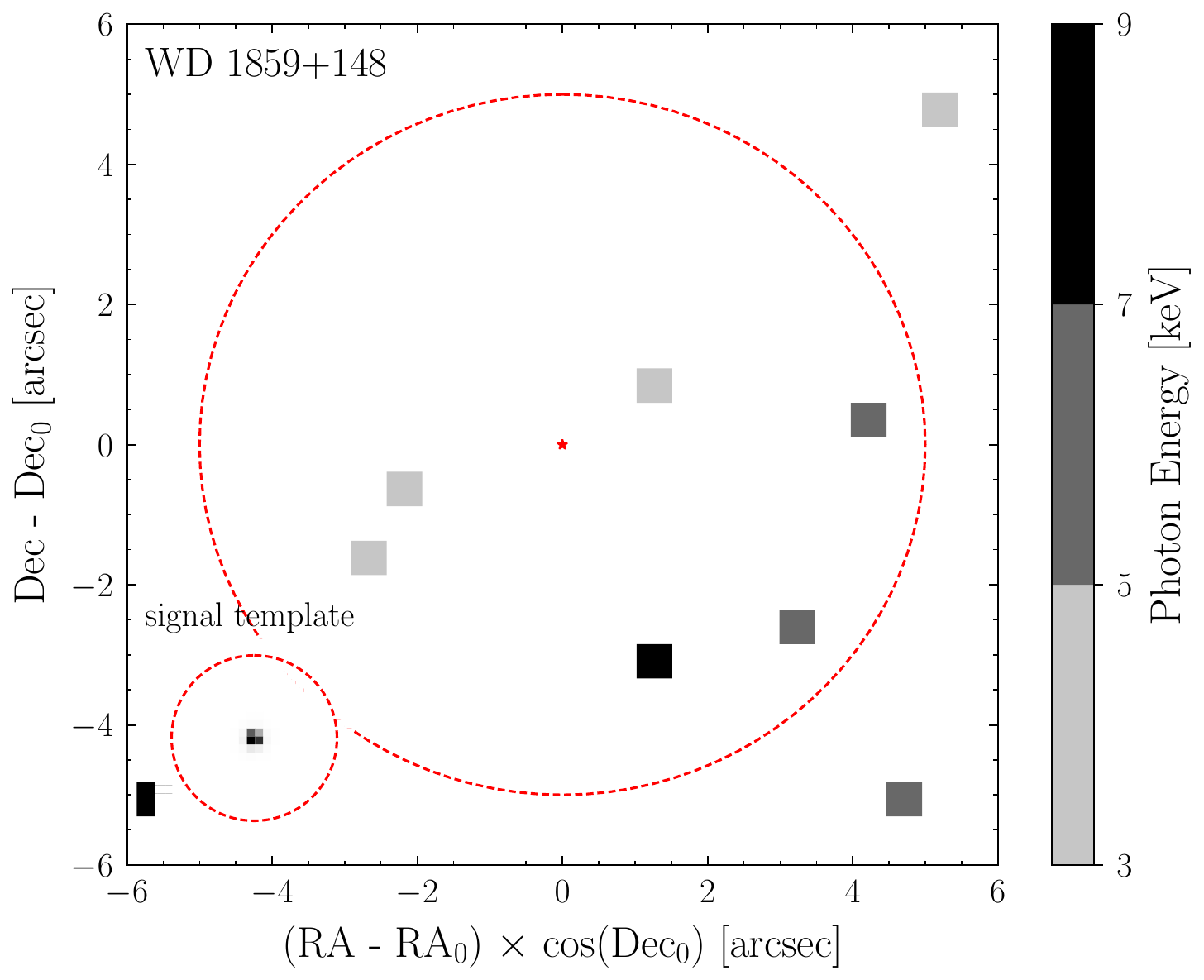}
\includegraphics[width=0.49\textwidth]{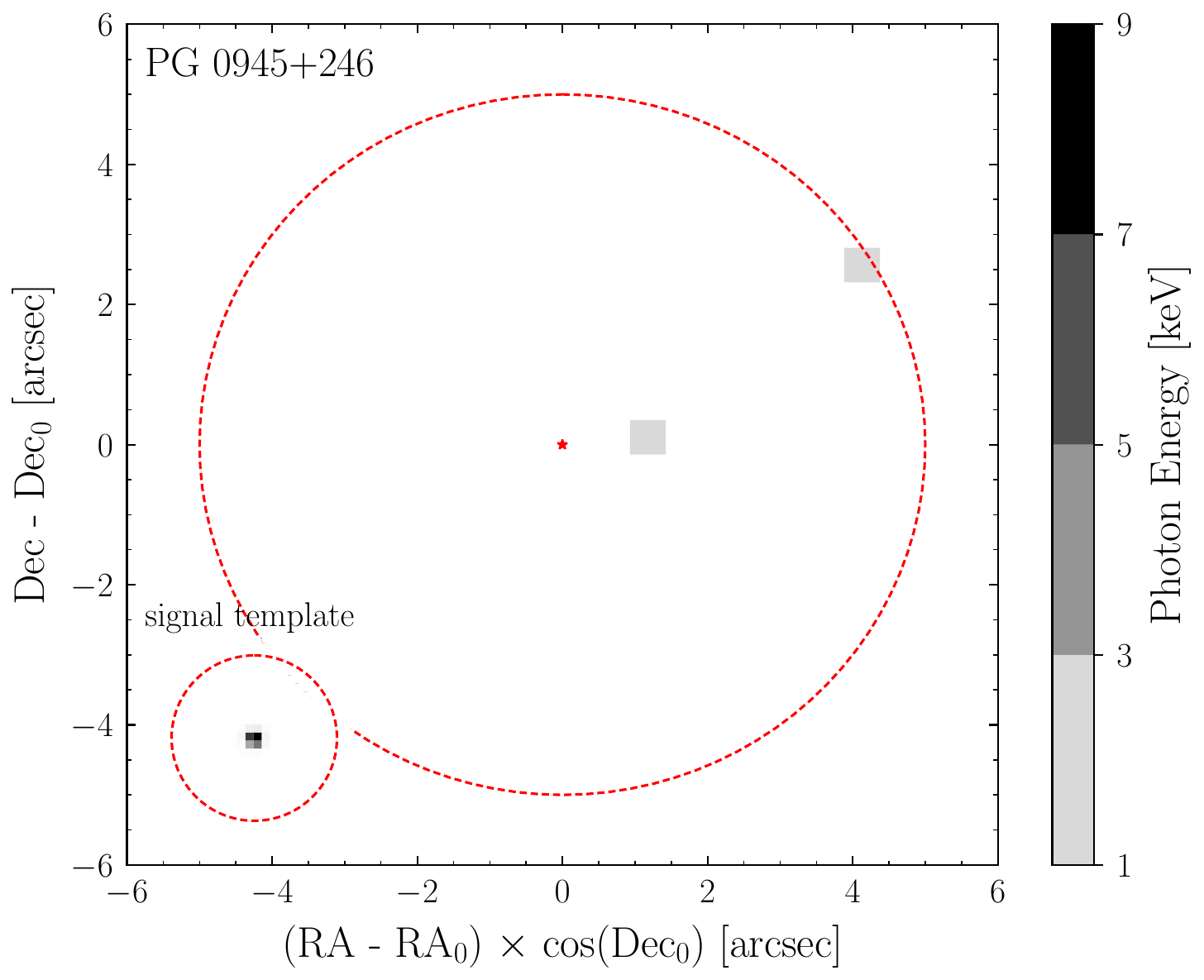}
\caption{(Left panel) The binned counts over the analysis range 3 -- 9 keV for WD 1859+148. No counts are observed within a PSF of the source. The dashed circle indicates the ROI used in our analysis. A maximum of one count was observed in each energy bin and pixel. The energy-averaged signal template is shown in inset, colored by intensity. (Right panel) Similarly, but for PG 0945+246 over the 1 -- 9 keV energy range.}
\label{fig:data_maps}
\end{figure*}

\begin{figure}[!htb]
\centering
\includegraphics[width=0.45\textwidth]{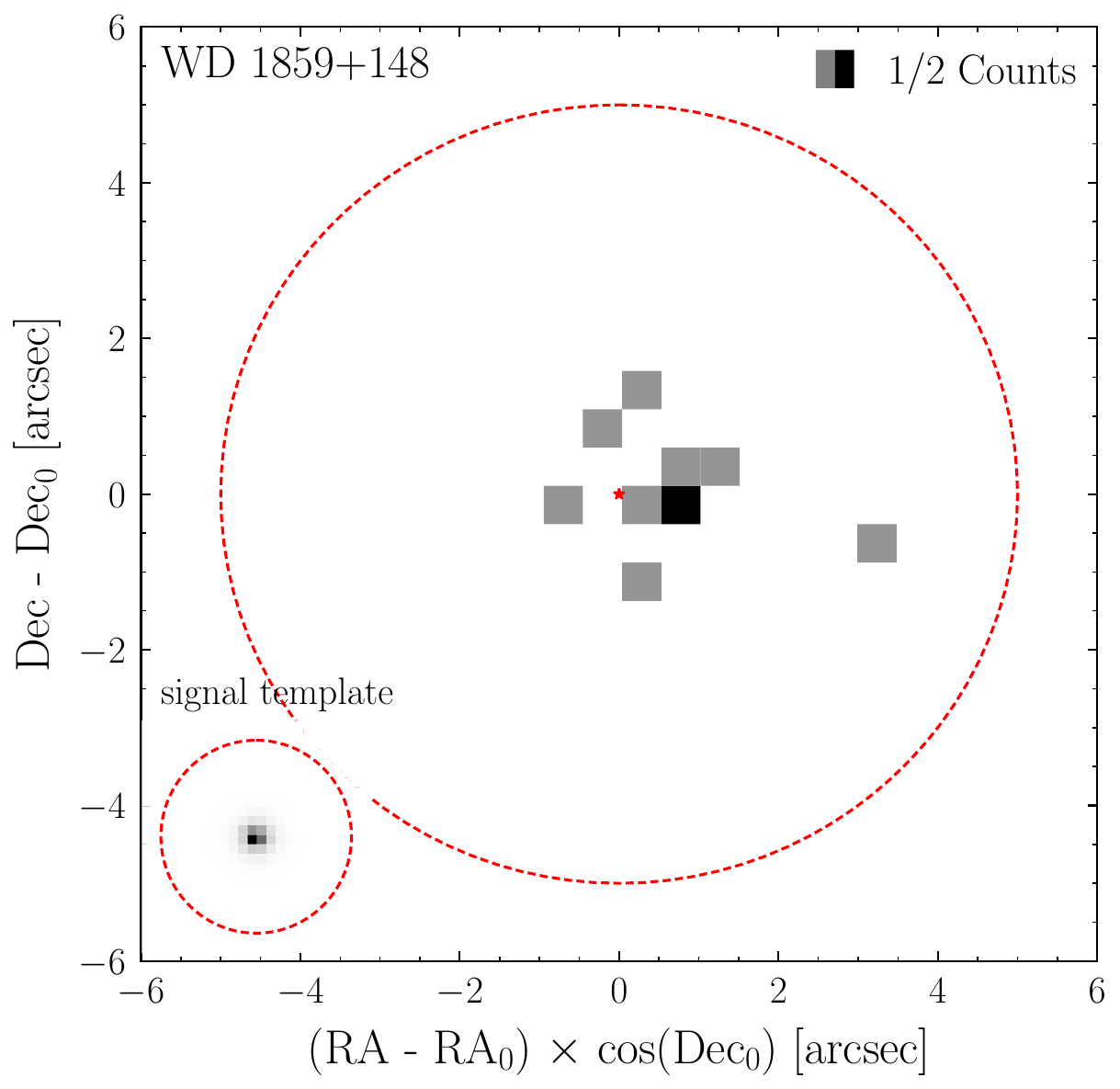}
\vspace{-0.4cm}
\caption{The same as Fig.~\ref{fig:data_maps}, but for the 1 -- 3 keV bin in the WD 1859+148 analysis. A clear excess is visible at the location of the source. Note that the coloring indicates the total number of counts. The signal template is broader than in other bins due to contamination on the optical blocking filter. }
\label{fig:excess_map}
\end{figure}

In Figs.~\ref{fig:data_maps} and ~\ref{fig:excess_map} we illustrate the pixelated and stacked data for both targets, along with the PSF templates used in the analysis described below.  An excess of low-energy X-rays below 3 keV is noticeable in WD 1859+148 data near the source, while PG 0945+246 shows no signs of X-ray emission above 1 keV.

\begin{figure}[!htb]
\centering
\includegraphics[width=0.49\textwidth]{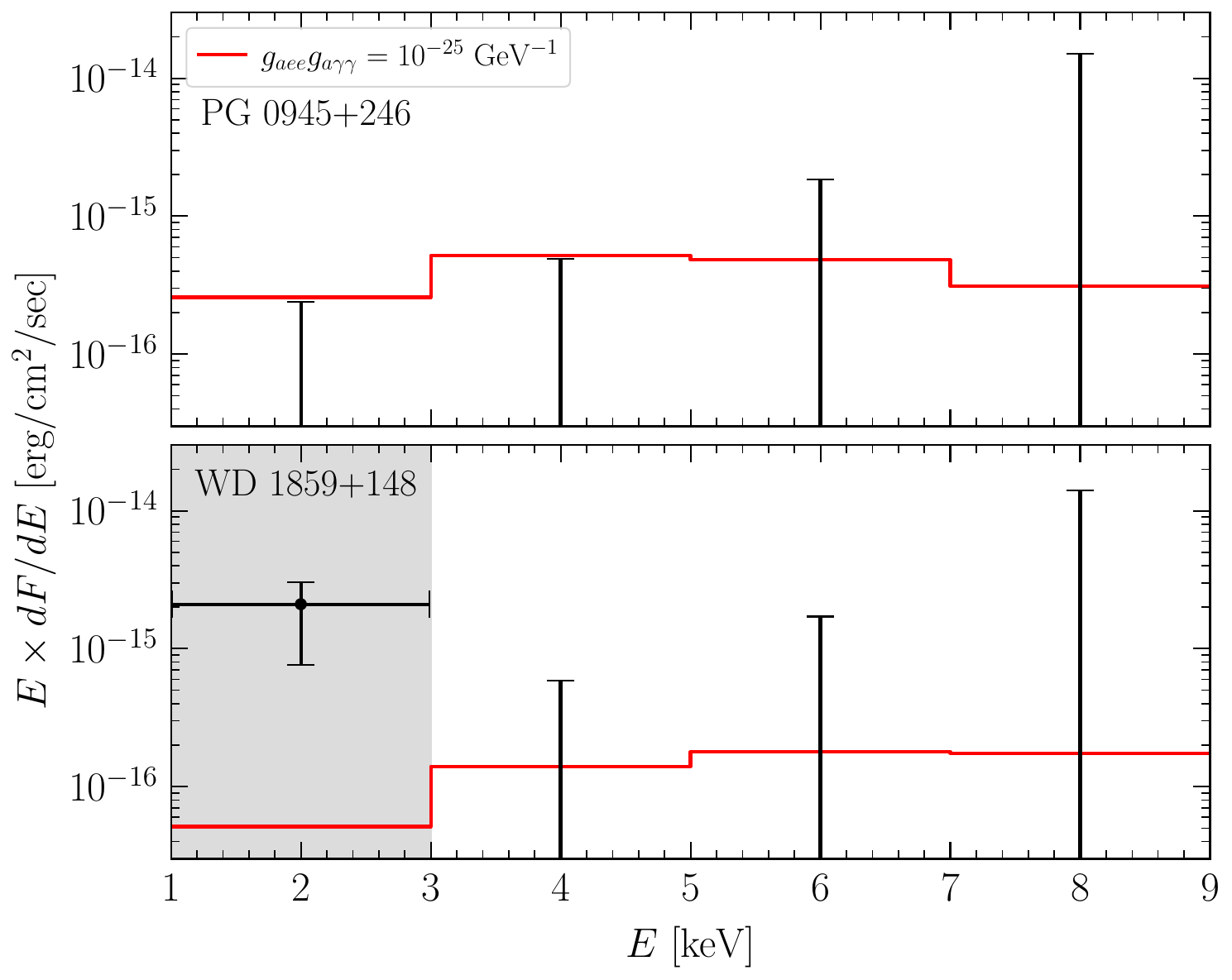}
\vspace{-0.4cm}
\caption{The energy spectrum across our four energy bins from the analysis of \textit{Chandra} data of MWDs PG 0945+246 and WD 1859+148. For both MWDs, the best-fit fluxes are consistent with zero, with the 68\% containment intervals in each energy bin shown. We additionally overlay predicted axion-induced signals with the indicated couplings in the massless limit. We note that for WD 1859+148, we do not analyze the 1 -- 3 keV bin due to observed non-axion astrophysical emission, as discussed in the main text.}
\label{fig:spectra_data}
\end{figure}

We analyze the stacked data for each target independently in each energy bin, through the procedure described below, in order to obtain the best-fit spectra for the MWDs shown in Fig.~\ref{fig:spectra_data}.
For a given MWD and in a given energy bin, labeled by $i$, we analyze the pixelated data $\boldsymbol{d} = \{n_{i,j}\}$ for each MWD, where $n_{i, j}$ is the number of counts in each energy bin $i$ and pixel $j$. We implement a spatial template fit through a joint Poisson likelihood
\begin{equation}
    p_i(\boldsymbol{d} | \mathcal{M}, \boldsymbol{\theta}) = \prod_{j=1}^{N_{\rm pix}} \frac{\mu_{i,j} (\boldsymbol{\theta})^{n_{i,j}} e^{-\mu_{i,j}(\boldsymbol{\theta})}}{n_{i,j}!} \,,
\label{eq:likelihood}
\end{equation}
where $\mathcal{M}$ indicates the combined signal and background model parameterized as $\boldsymbol{\theta} = \{ A_{\rm bkg}, A_{\rm sig}\}$, with $N_{\rm pix}$ denoting the number of spatial pixels.  We include all pixels within $5\farcs0$ of the source center in our analysis region of interest (ROI). 
The number of counts predicted by our combined model is denoted as $\mu_{i,j}(\boldsymbol{\theta})$, while the background parameter $A_{\rm bkg}$ indicates a normalization parameter that re-scales the background counts spatial template. The background can arise from astrophysical sources and X-ray fluorescence from within the instrument. There are no astrophysical point sources in either of our ROIs, so we model the background in both cases as spatially flat. For our signal model, the number of predicted counts in each energy bin is determined by $A_{\rm sig}$, which is the signal flux in units of cts/cm$^2$/s.  The signal flux is mapped to predicted counts using the instrument response. The signal spatial template is given by the energy-dependent {\it Chandra} PSF at the location of the source.  

We profile over the nuisance parameter $A_{\rm bkg}$, which is independent in each energy bin, in order to construct the profile likelihood for the source flux $A_{\rm sig}$ associated with the MWD in each energy bin (see, {\it e.g.},~\cite{Safdi:2022xkm} for details of the statistical procedure).  In Fig.~\ref{fig:spectra_data} we show the best-fit flux and associated $1\sigma$ confidence intervals for each of the MWDs across all of the energy bins considered.  Note that we illustrate the results in terms of $E dF / dE \equiv E_{\rm cent}^2 A_{\rm sig} / dE$, where $E_{\rm cent}$ is the central energy of the energy bin and $dE$ is the energy bin width.  

All energy bins are consistent with zero signal counts for both MWDs except for the 1 -- 3 keV energy bin for WD 1859+148.  
In particular, we find $2.6\sigma$ evidence for emission in the 1 -- 3 keV energy range at the location of the WD, with integrated flux $(1.2 \pm 0.4) \times 10^{-15}$ erg/cm$^2$/s. However, the emission cannot be associated with axion-induced X-rays, because we observe no counts in the hard X-rays ($>3$ keV), which would be predicted for the axion model as discussed below. Indeed, to explain the observed data with axion emission the core temperature of the MWD would need to be below a keV. This scenario is strongly disfavored by our analysis, described below, and furthermore the required axion couplings in that scenario are excluded. For this reason, we mask the 1 -- 3 keV bin from our analysis of WD 1859+148. The emission is interpreted as arising from magnetospheric radiation in Ref.~\cite{Bamba:2024nwd}.

\section{Interpretation of results for the axion model}

The axion emission from the interiors of our MWDs primarily comes from electron-nuclei bremsstrahlung scattering of the form $e \,+\, (A, Z) \to e \,+ \,(A, Z)\, + \,a$\,, with $A$ the nuclei mass number and $Z$ the atomic number. Furthermore, since our physical system is a WD core, the electrons are strongly degenerate with temperature $T \ll p_F$ much smaller than the Fermi momentum $p_F$. Consequently, the axion emissivity is thermal and calculated as~\cite{Raffelt:1990yz, Nakagawa:1987pga}
\begin{equation}
    \frac{d\varepsilon_a}{d\omega} = \frac{\alpha_{\rm EM}^2 g^2_{aee}}{4\pi^3 m_e^2} \frac{\omega^3}{e^{\omega/T} - 1} \sum_s \frac{Z_s^2 \rho_s F_s}{A_s u} \,.
\label{eq:dedw}
\end{equation}
The sum is over nuclei species $s$ present in the WD plasma, while $Z_s$ is the atomic number, $A_s$ is the mass number, $\rho_s$ is the mass density, and $u \simeq 931.5$ MeV is the atomic mass unit. Medium effects are taken into account by the species-dependent, dimensionless factors $F_s$; these effects include the screening of the electric field and interference between different scattering sites~\cite{Ichimaru:1982zz}. We use the empirical fitting functions provided in~\cite{Nakagawa:1988rhp} for strongly-coupled plasmas to calculate $F_s$. 

To calculate the quantity~\eqref{eq:dedw}, we utilize WD models simulated using the Modules for Experiments in Stellar Astrophysics (MESA) code package~\cite{2011ApJS..192....3P, 2013ApJS..208....4P} in conjunction with updated \textit{Gaia} luminosity measurements, broadly following the techniques used in~\cite{Dessert:2021bkv}. Stellar profiles of WD densities and compositions are extracted through MESA, which can simulate and evolve WDs of various masses throughout their lifetimes. These simulations account for important phenomena relevant to emulating realistic WD interior plasmas, including effects from ionic correlations and crystallization in the core. On the other hand, the core temperatures of our MWDs are primarily inferred from \textit{Gaia} photometric data; by fitting WD cooling sequences~\cite{Camisassa:2022pet} to predicted \textit{Gaia} DR2 band magnitudes~\cite{GaiaDR2}, one can estimate the core temperature of a WD, along with, for example, its cooling age. We use this information to then select our fiducial MESA model, which finally allows us to completely determine~\eqref{eq:dedw} for a given WD (see App.~\ref{app:wdmodeling} for further details).

\begin{figure}[!htb]
\centering
\begin{minipage}{\columnwidth}
\includegraphics[width=\columnwidth]{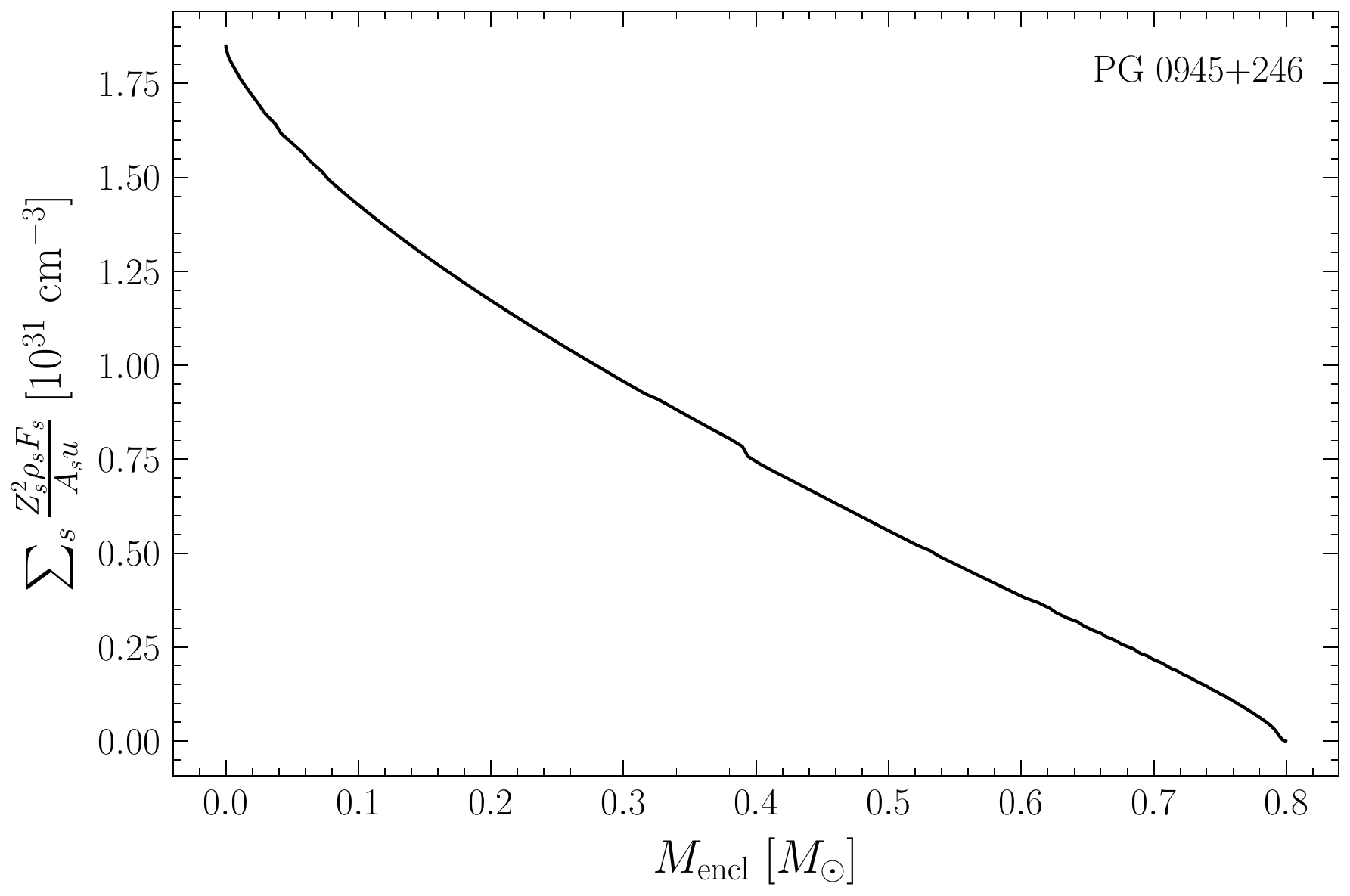}
\end{minipage}
\caption{The profile of the sum in~\eqref{eq:dedw} for our fiducial MESA model for PG 0945+246. This sum is an important intermediate quantity for understanding the total axion emissivity of our MWDs.}
\label{fig:intermediate}
\end{figure}

The procedure described above leads us to adopt a fiducial WD model with core temperature $T_c \approx 1.50$ $(2.20)$ keV, mass $0.8$ $M_{\odot}$ ($1.33$ $M_{\odot}$), and composition C/O (O/Ne) for PG 0945+246 (WD 1859+148) (see App.~\ref{app:wdmodeling}). With these parameters, the total axion luminosity for our WDs is then found by integrating \eqref{eq:dedw} over the WD cores.  In Fig.~\ref{fig:intermediate} we illustrate the combination of WD-specific quantities appearing after $\sum_s$ in~\eqref{eq:dedw} for our fiducial model for PG 0945+246. 

After being produced in the WD interiors, the axions stream out and convert to X-rays in the MWD magnetic fields. The conversion probability $p_{a \to \gamma}$ is numerically calculated for arbitrary magnetic field configurations and axion masses $m_a$ by solving the axion-photon mixing equations, importantly incorporating the Euler-Heisenberg Lagrangian terms which suppresses the mixing by modifying the photon propagation in the strong magnetic fields~\cite{Raffelt:1987im}. We follow the prescription in~\cite{Dessert:2019sgw} and calculate the conversion probabilities assuming a magnetic dipole model of strengths $\sim$670 MG and $\sim$800 MG for PG 0945+246 and WD 1859+148, respectively~\cite{Ferrario:2015oda, Caiazzo:2021xkk} (although see~\cite{Dessert:2019sgw} for possible enhancements due to alternate models).  Note that these magnetic field measurements are inferred from Zeeman shifts in optical spectral data due to absorption lines in the MWD atmospheres in the presence of the strong surface fields. We neglect model-dependent mass mixing between axion states that could reduce the conversion probability~\cite{Chadha-Day:2023wub}.

\begin{figure}[!htb]
\centering
\begin{minipage}{\columnwidth}
\includegraphics[width=\columnwidth]{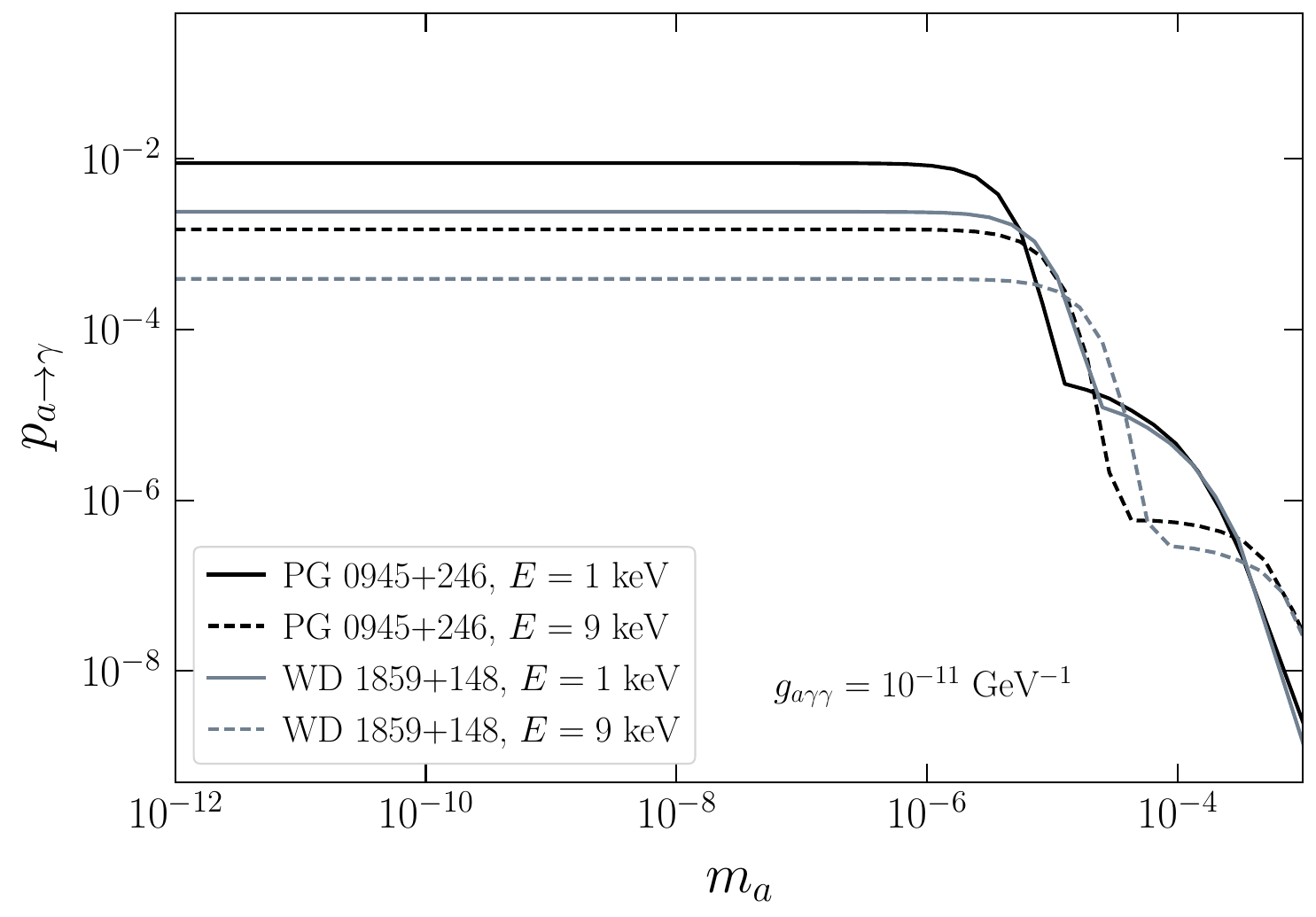}
\end{minipage}
\caption{Axion-to-photon conversion probabilities $p_{a\to \gamma}$ assuming axions of energy 1 keV (solid) and 9 keV (dashed), for both of our MWDs, as a function of axion mass $m_a$, at the indicated axion-photon coupling.}
\label{fig:conv}
\end{figure}

In Fig.~\ref{fig:conv} we illustrate the axion-to-photon conversion probabilities for 1 keV and 9 keV axions for each of our target MWDs as a function of the putative axion mass $m_a$.
The conversion probabilities are roughly independent of energy at low $m_a$ and are independent of $m_a$ below a critical value. The critical value of $m_a$ may be estimated by asking when the phase shift $\delta \phi \sim m_a^2 / (2 E) R_{\rm wd}$ between the axion and photon, propagating with energy $E$ over a distance of order the MWD radius $R_{\rm wd}$, becomes larger than unity. This suggests that the axion-to-photon conversion probability should be, for example, independent of $m_a$ for $m_a \lesssim 10^{-5}$ eV for $E = 1$ keV, matching Fig.~\ref{fig:conv}.  Note that at low $m_a$, $p_{a\to\gamma} \propto B^{2/5}$~\cite{Dessert:2019sgw} due to the Euler-Heisenberg term, which modifies the dispersion relation of the photon.  Given that the flux scales with the axion couplings as $|g_{aee} \times g_{a\gamma\gamma}|^2$, this implies that our sensitivity to $|g_{aee} \times g_{a\gamma\gamma}|$ scales weakly with the assumed dipole magnetic field strength as $B^{-1/5}$, indicating that our choice of $B$ is a subdominant source of uncertainty on our final result.

Once the axions convert to photons, they can be detected with \textit{Chandra} after traveling a distance $d$ between the source and Earth. We take $d = 36.20$ $(41.40)$ pc for PG 0945+246 (WD 1859+148) inferred from \textit{Gaia}~\cite{2020arXiv201201533G}. We note that we take the central values measured from \textit{Gaia} for $d$, because the uncertainties on the distances only leads to a $\sim$$0.1$\% uncertainty on the axion-induced flux. 
In Fig.~\ref{fig:spectra_data} we illustrate example signal fluxes for $m_a \ll 10^{-5}$ eV and $|\gaee \times \gagg| = 10^{-25}$ GeV$^{-1}$ for both of our targets. (See also App. Fig.~\ref{fig:PG_spectra_cts} and App. Fig.~\ref{fig:ZTF_spectra_cts} for illustrations of how these spectra map to counts in each energy bin.)  Note, in particular, that the low-energy excess in WD 1859+148 is clearly inconsistent with an axion origin because the axion model has too hard of a spectrum. Explaining the excess with an axion model would require us to adopt a lower WD core temperature, which is inconsistent with observations and which would also require such large $|\gaee \times \gagg|$ values that they would already be excluded by other probes.  We thus exclude the 1 -- 3 keV energy bin in our analysis of WD 1859+148.

For each MWD, we constrain the signal model by using a joint likelihood 
\begin{equation}
    p(\boldsymbol{d} | \mathcal{M}_{\rm axion}, \boldsymbol{\theta}) = \prod_{i =1}^{N_{\rm bins}} p_i(\boldsymbol{d} | \mathcal{M}, \boldsymbol{\theta})\,,
\label{eq:axion_likelihood}
\end{equation}
where $\mathcal{M}_{\rm axion}$ indicates the combined axion signal and background model, $N_{\rm bins}$ is the number of energy bins, and the $p_i$ are the likelihoods given in~\eqref{eq:likelihood}.  The axion model has parameters $\boldsymbol{\theta} = \{ \boldsymbol{A}_{\rm bkg}, |\gaee \times \gagg|, m_a\}$, where the vector $\boldsymbol{A}_{\rm bkg}$ indicates a background normalization parameter in each energy bin. The signal flux vector in each energy bin is determined by the modeling procedure described above, so that we may write $\boldsymbol{A}_{\rm sig}(\{ \gaee\gagg, m_a\})$. 

Given our likelihood~\eqref{eq:axion_likelihood}, we construct the profile likelihood for $|\gaee \times \gagg|$ at fixed $m_a$ by profiling over $\boldsymbol{A}_{\rm bkg}$ (see, {\it e.g.},~\cite{Cowan:2010js,Safdi:2022xkm}). Because we are in the low-counts limit, the 95\% upper limits on $|\gaee \times \gagg|$ are inferred by Monte Carlo simulations of the signal and null hypotheses (as opposed to utilizing Wilks' theorem, see~\cite{Cowan:2010js}). We also power-constrain our limits to account for the possibility of under fluctuations (see~\cite{Cowan:2011an}), although this is ultimately not necessary in practice given that the data show no under fluctuations.   

We find no evidence for axions, with the best-fit coupling combination being zero across both MWDs. We instead set one-sided 95\% upper limits on $|\gagg \times \gaee|$ as a function of the axion mass for each of our MWDs, illustrated in Fig.~\ref{fig:gaeegagg}. In the massless limit, our 95\% upper limits are $|\gagg\gaee| \lesssim 1.54 \times 10^{-25}$ ($3.54 \times 10^{-25}$) GeV$^{-1}$ for PG 0945+246 (WD 1859+148). The joint limit (black in Fig.~\ref{fig:gaeegagg}), which combines the profile likelihoods from our MWDs as well as that from RE J0317-853, gives $|\gagg\gaee| \lesssim 1.01 \times 10^{-25}$ GeV$^{-1}$ at low axion masses.  In Fig.~\ref{fig:gaeegagg} we also illustrate the expected 95\% power-constrained upper limits, at 1$\sigma$ and 2$\sigma$ containment, under the null hypothesis as determined through Monte Carlo simulations.  These Monte Carlo simulations use the best-fit parameter vector $\boldsymbol{A}_{\rm bkg}$ from the fit under the null hypothesis. Our upper limits are consistent with the expectations under the null hypothesis.  Note that while our upper limit from PG 0945+246 is nearly identical to that from RE J0317-853~\cite{Dessert:2021bkv}, the upper limit from WD 1859+148 is subdominant. This is due mostly to our exclusion of the 1 -- 3 keV energy bin in that analysis. 
 Indeed, in Fig.~\ref{fig:gagggaee_oldEA} we show what the 95\% upper limit would have been had there not been a 1 -- 3 keV excess in the WD 1859+148 data, in which case it is comparable to that from RE J0317-853. We also illustrate in that figure that the sensitivity of PG 0945+246 would have passed that of RE J0317-853 had it not been for the instrumental degradation between 2020 and 2022/2023.   

\section{Discussion}

In this work we set some of the strongest constraints to-date on the coupling constant combination $|\gaee \times \gagg|$ for $m_a \lesssim 10^{-5}$ eV using dedicated X-ray observations with the {\it Chandra} observatory towards two nearby MWDs. One of the MWDs, WD 1859+148, shows an excess of low-energy X-rays below 3 keV, though the flux appears too soft to have an axion origin. 

Given the strong theoretical motivations for ultralight axions with weak couplings to ordinary matter, it is interesting to consider how more sensitive versions of the search presented in this work could be pursued in the future. Our work strongly suggests that given the degradation of {\it Chandra} over the past few years it is unlikely that more promising observations could be made towards other nearby MWD targets with that instrument. On the other hand, a number of {\it Chandra} followup missions have been proposed. For example, as discussed in~\cite{Dessert:2021bkv} the proposed Lynx X-ray telescope~\cite{LynxTeam:2018usc} could improve the sensitivity to $|\gaee \times \gagg|$ by well over an order of magnitude by observations of the same targets studied in this work due largely to the substantially larger effective area of Lynx relative to Chandra. By contrast, the proposed NewAthena telescope~\cite{Nandra:2013jka} could reach sensitivities to axion-induced X-rays approximately a factor of three stronger than those derived in this work. Lynx could launch in the 2030s, while NewAthena is planned to launch in 2037.

A more promising approach for probing low-mass axions relative to that presented in this work could be to perform X-ray observations of nearby galaxies as in~\cite{Ning:2024eky} and study axion production in {\it e.g.} WD populations. The axions could convert to photons in this case both off of the MWD magnetic fields but also on the galactic fields, though the galactic fields would not efficiently convert axions with masses $m_a \sim 10^{-5}$ eV as in this work because of their larger extensions.  Such population studies could also account for axion-electron production in other stellar systems, such as massive stars.  We leave such studies to future work.

\begin{acknowledgments}
{\it 
We thank J. Berryman for collaboration in the early stages of this work. We thank J. Benabou, H. Guenther, and Y. Park for useful discussions. B.R.S was supported in part by the DOE Early Career Grant DESC0019225 and in part by the DOE award DE-SC0025293. The work of O.N. is supported in part by the NSF Graduate Research Fellowship Program under Grant DGE2146752. Support for this work was provided by the National Aeronautics and Space Administration through {\it Chandra} Award Number GO0-21013X issued by the {\it Chandra} X-ray Center (CXC), which is operated by the Smithsonian Astrophysical Observatory for and on behalf of the National Aeronautics Space Administration under contract NAS8-03060. The scientific results reported in this article are based to a significant degree on observations made by the {\it Chandra} X-ray Observatory. This research has made use of software provided by the CXC in the application package CIAO.
}

\end{acknowledgments}

\appendix

\section{Supplementary Figures}
\label{app:suppfig}

In this section we illustrate several supplementary figures relevant to our methods and analysis, which are referenced in the main text.

\begin{figure}[!htb]
\centering
\includegraphics[width=\columnwidth]{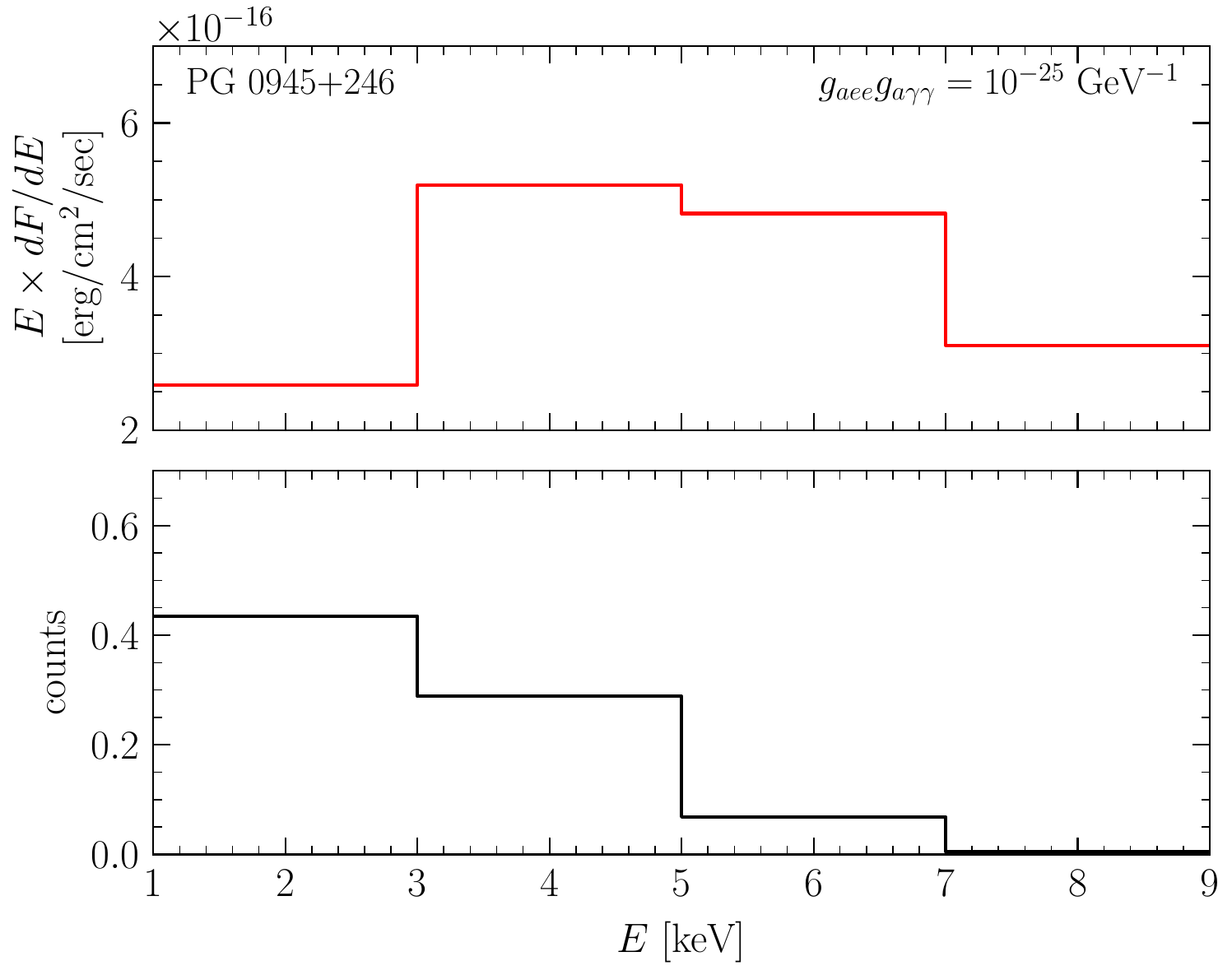}
\caption{\label{fig:PG_spectra_cts} (Top) The predicted axion-induced energy spectrum from the MWD PG 0945+246, with the indicated coupling, for $m_a \ll 10^{-5}$ eV. (Bottom) The predicted forward-modeled counts from PG 0945+246, taking into account factors such as the \textit{Chandra} effective area and total exposure time.}
\end{figure}

\begin{figure}[!htb]
\centering
\includegraphics[width=\columnwidth]{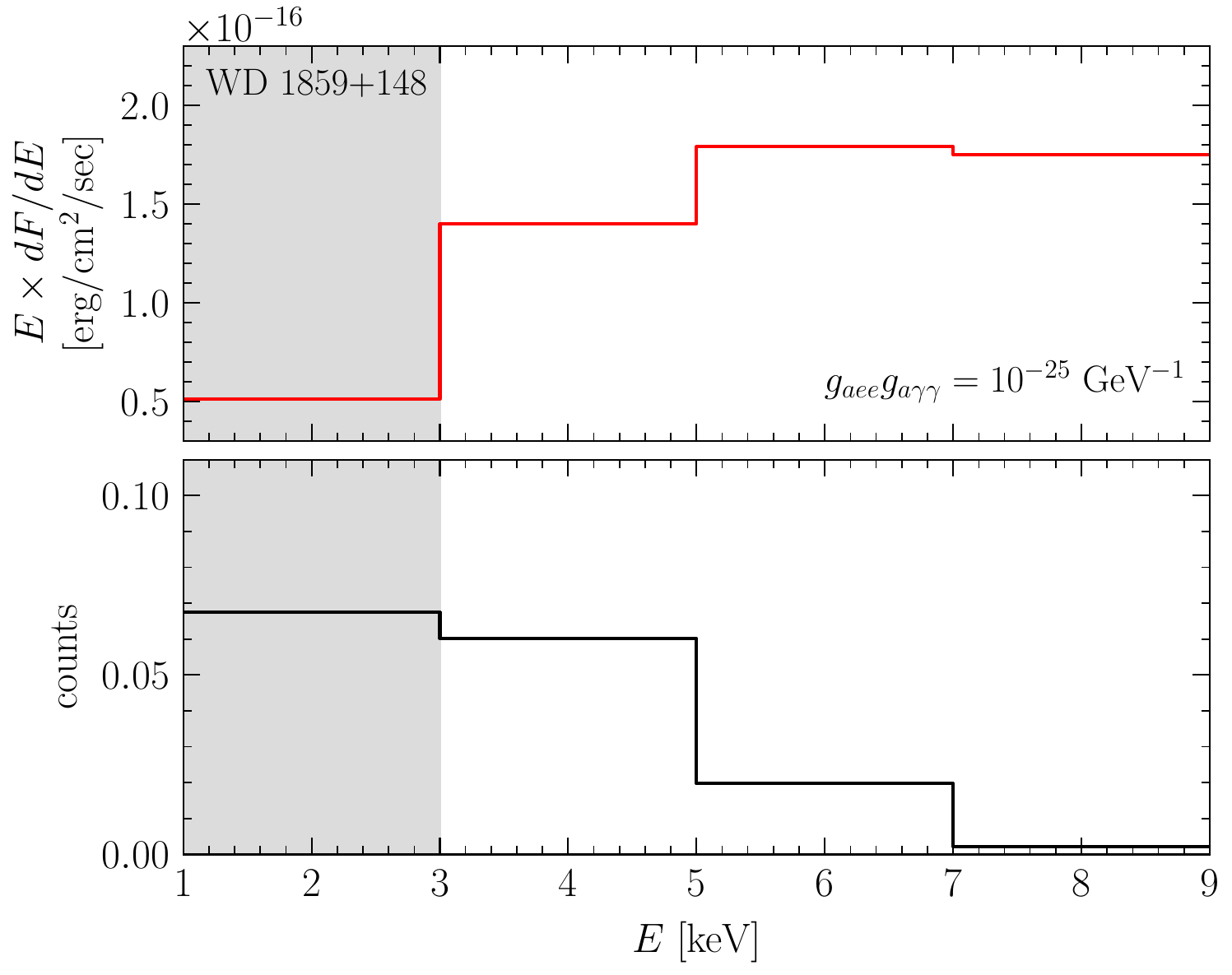}
\caption{\label{fig:ZTF_spectra_cts}The same as Fig.~\ref{fig:PG_spectra_cts} but for MWD WD 1859+148. Note that we do not analyze bins below 3 keV due to the observation of non-axion astrophysical emission, discussed in the main text.}
\end{figure}

\begin{figure}[H]
\centering
\includegraphics[width=\columnwidth]{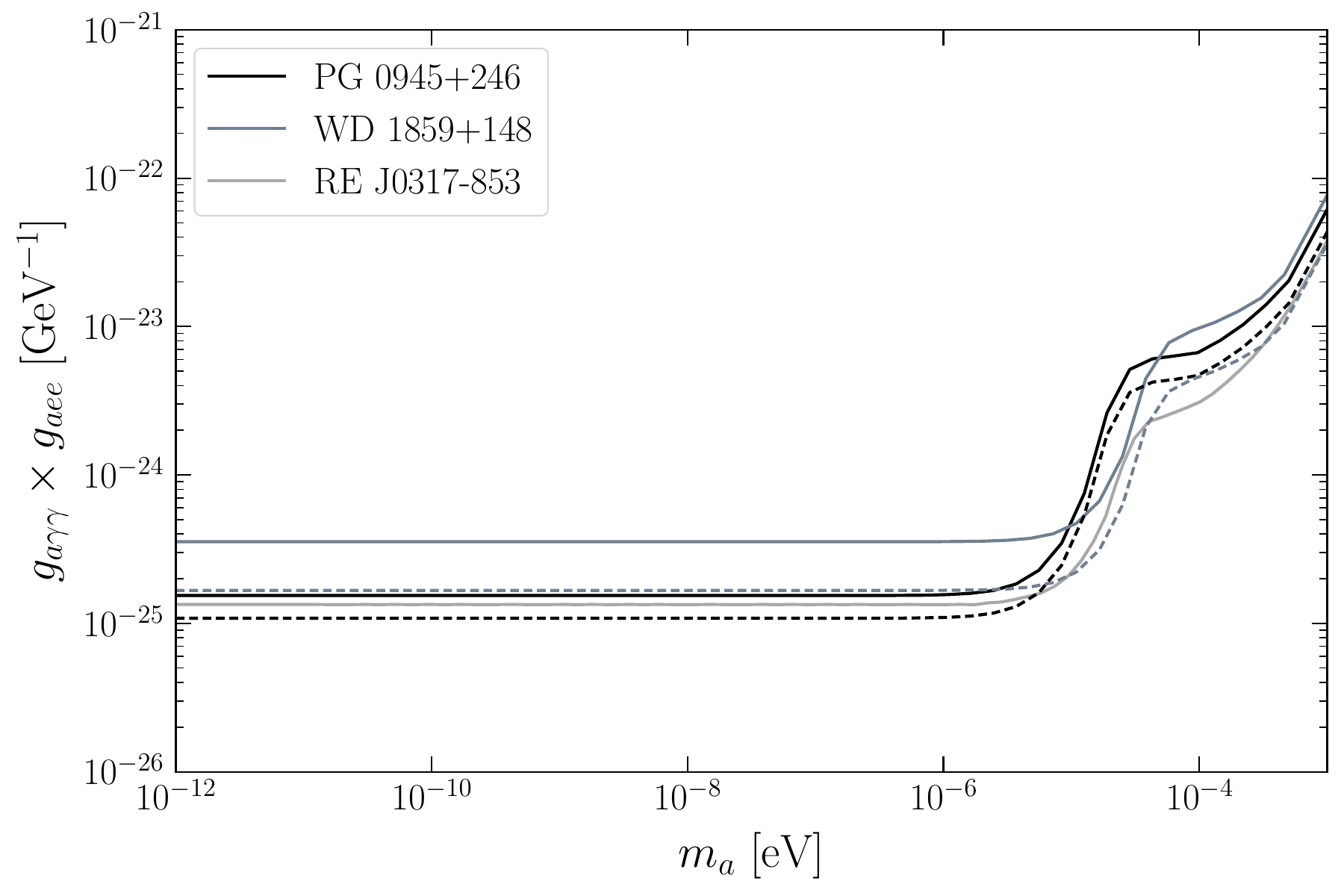}
\caption{\label{fig:gagggaee_oldEA} A comparison of the expected 95\% upper limits on $\gagg \gaee$ from PG 0945+246 and WD 1859+148 using the corresponding contemporary \textit{Chandra} effective areas (solid), as in Fig.~\ref{fig:gaeegagg}, and the analogous 95\% upper limits but assuming the effective area corresponding to the observational epoch of RE J0317-853~\cite{Dessert:2021bkv} (dashed). We also assume no excess counts in the first energy bin for the WD 1859+148 analysis in making the dashed projection. As discussed in the main text, and apparent in Fig.~\ref{fig:eff_areas}, the reduction of the effective area over the past few years and the observation of background counts reduce the strength of our final results.}
\end{figure}

\section{Data reduction and calibration}
\label{app:data}

The data from the {\it Chandra} ACIS-I \texttt{Timed Exposure} observations of WD 1859+148 are reduced as follows. For the data reduction process, we use the {\it Chandra} Interactive Analysis of Observations (CIAO)~\cite{2006SPIE.6270E..1VF} version 4.11. We reprocess each observation with the CIAO task \texttt{chandra\_repro}, which produces an events file filtered for flares and updated for the most recent calibration. We create counts and exposure images (units [cm$^2$s]) with pixel sizes of $0\farcs492$ and in 2 keV energy bins for energies between 0 and 9 keV with \texttt{flux\_image}. We then account for the proper motion of the WD. WD 1859+148 was observed by {\it Gaia} in the DR3 with location ${\rm RA} \approx 19^\circ \, 01' \, 32\farcs85$, ${\rm DEC} \approx 14^\circ \, 58' 08\farcs34$ at the reference epoch of J2016.0~\cite{2020arXiv201201533G}.  We use the proper motion measurements from {\it Gaia} to infer the position during each observation (taken between December 9, 2022 and December 10, 2022). 

We process the data for PG 0945+246 identically, including accounting for proper motion using {\it Gaia} DR3 data to infer its position during the observation on December 16, 2023.

\section{White-Dwarf Modeling}
\label{app:wdmodeling}

In this section we detail and illustrate our modeling procedures for both MWDs WD 1859+148 and PG 0945+246. The complete modeling of the WD interiors are important for this analysis, as the expected axion luminosity spectra depends on the core temperature, the density profile, and the chemical abundance profiles throughout the WD core. Due to the high thermal conductivity of the degenerate matter in the WD core, we assume the core temperature is uniform throughout our MWDs, but allow the density and chemical abundances to vary. 

For PG 0945+246, following the formalism developed in ~\cite{Dessert:2021bkv}, we infer the core temperature and age of this MWD through photometric \textit{Gaia} observations and measurements in conjunction with WD cooling sequences~\cite{Camisassa:2022pet}, and then turn to MESA~\cite{2011ApJS..192....3P, 2013ApJS..208....4P} simulations of WDs with these properties to extract detailed density and chemical abundance profiles. From this analysis we find $T_c = 1.501 \pm 0.003$ keV, and a best-fit cooling age of $t \sim 0.37$ Gyr. Note that our inferred core temperature is consistent with the MESA simulation and other estimates, {e.g.} ones based on the surface temperature~\cite{Chabrier:2000ib}. Given the inferred core temperature and age of PG 0945+246, we use the MESA code \texttt{r23.05.1} to model the interior of this WD in order to procure density and chemical abundance profiles. MESA is a one-dimensional code that solves the equations of stellar structure and evolution. We simulate a MWD of the same mass as PG 0945+246, $0.8 \pm 0.01$ $M_{\odot}$~\cite{2024MNRAS.527.8687O}, using the default inlist suite to simulate WDs, \texttt{make\_c\_o\_wd}. We extract spatial density and chemical abundance profiles at $\sim 0.37$ Gyr which are used in our analysis.

\begin{figure}[!htb]
\centering
\begin{minipage}{\columnwidth}
\includegraphics[width=\columnwidth]{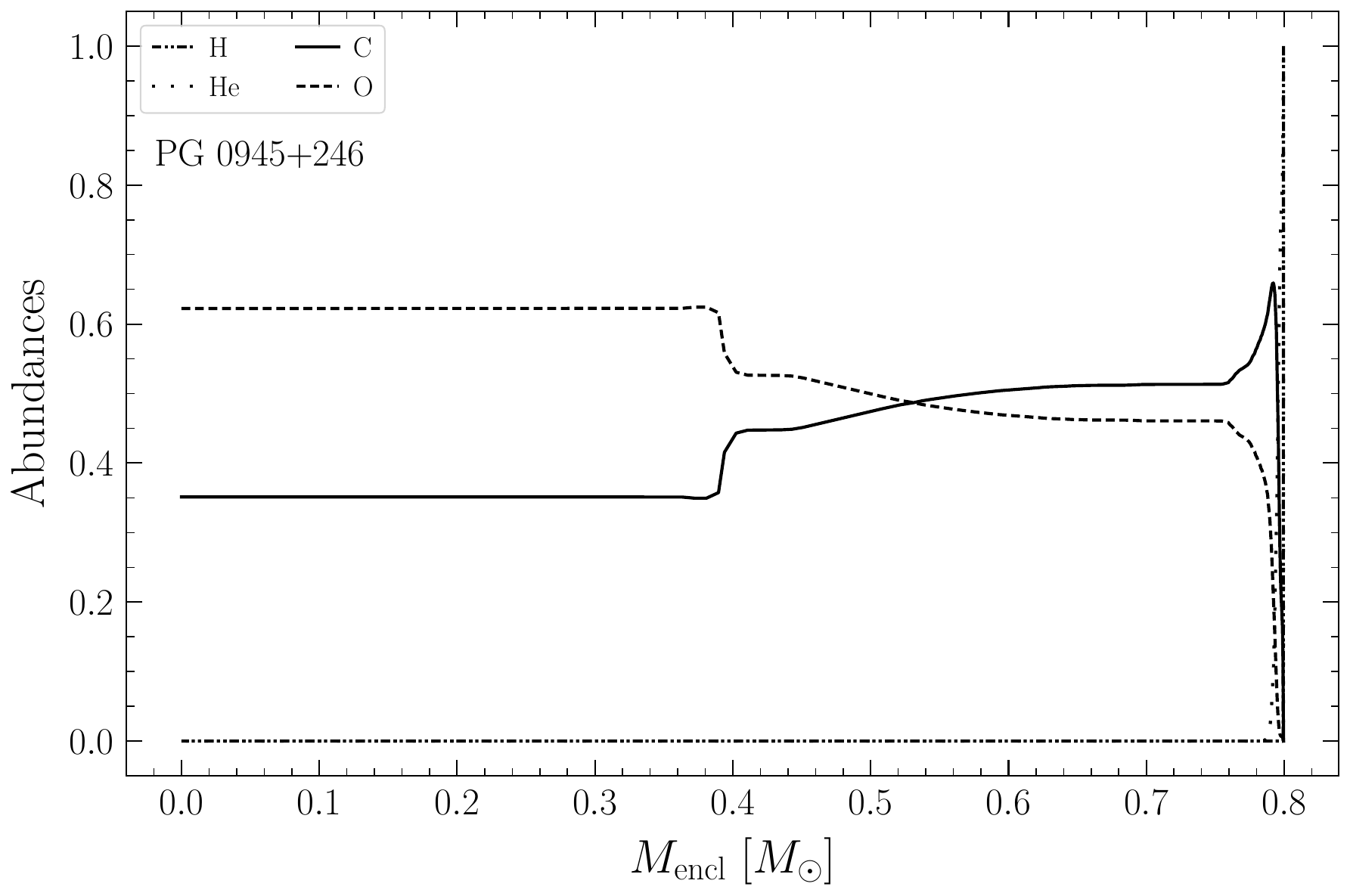}
\end{minipage}
\vspace{0.5cm}

\begin{minipage}{\columnwidth}
\includegraphics[width=\columnwidth]{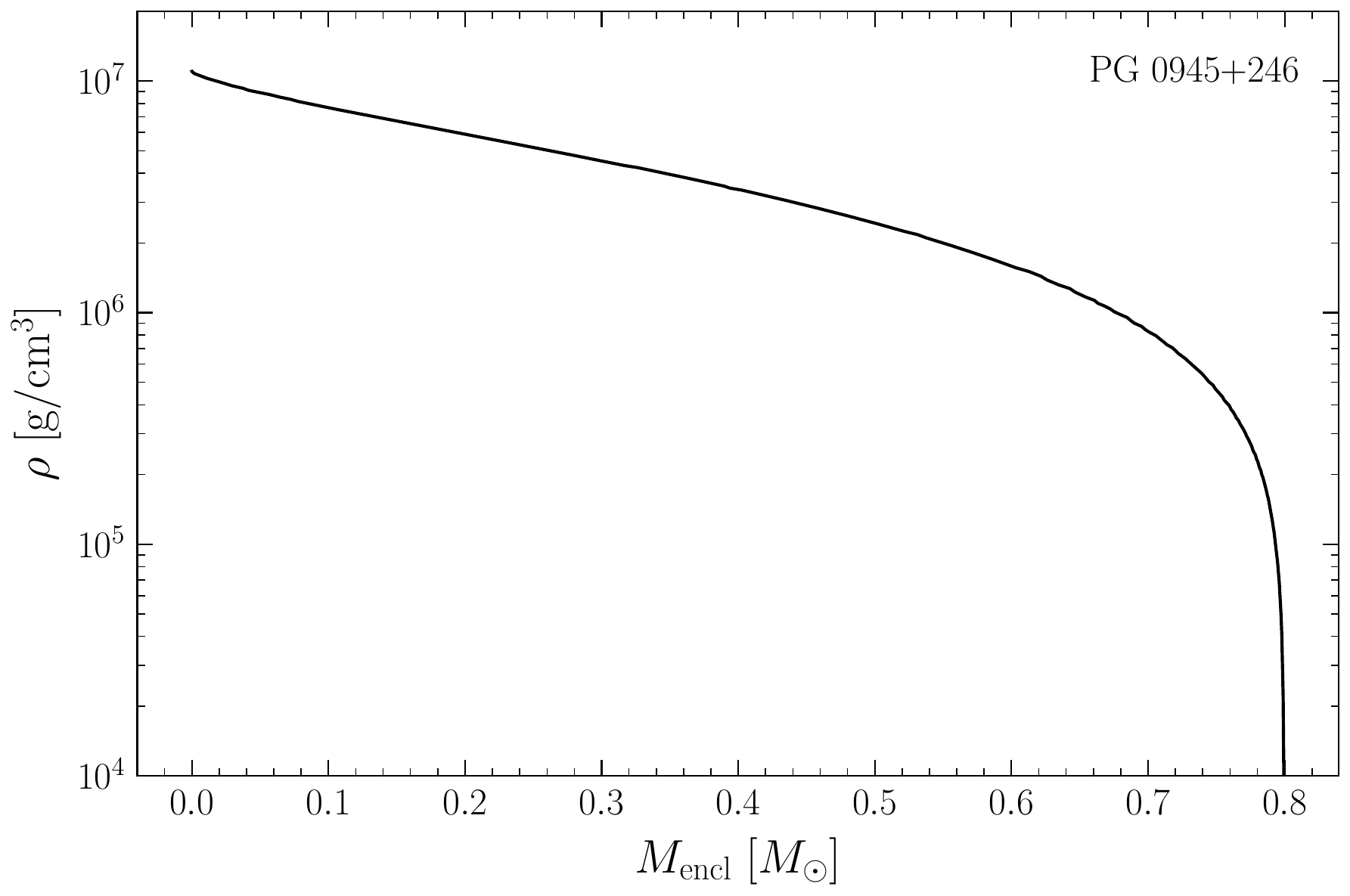}
\end{minipage} 
\caption{(Top) The hydrogen, helium, carbon, and oxygen chemical abundances from our fiducial MESA model for PG 0945+246. The x-axis is the enclosed mass coordinate. We note that these abundances are characteristic of a carbon-oxygen WD, which is what PG 0945+246 is expected to be. (Bottom) The density profile for the same model as a function of enclosed mass.}
\label{fig:PG_MESA}
\end{figure}

\begin{figure}[!htb]
\centering
\begin{minipage}{\columnwidth}
\includegraphics[width=\columnwidth]{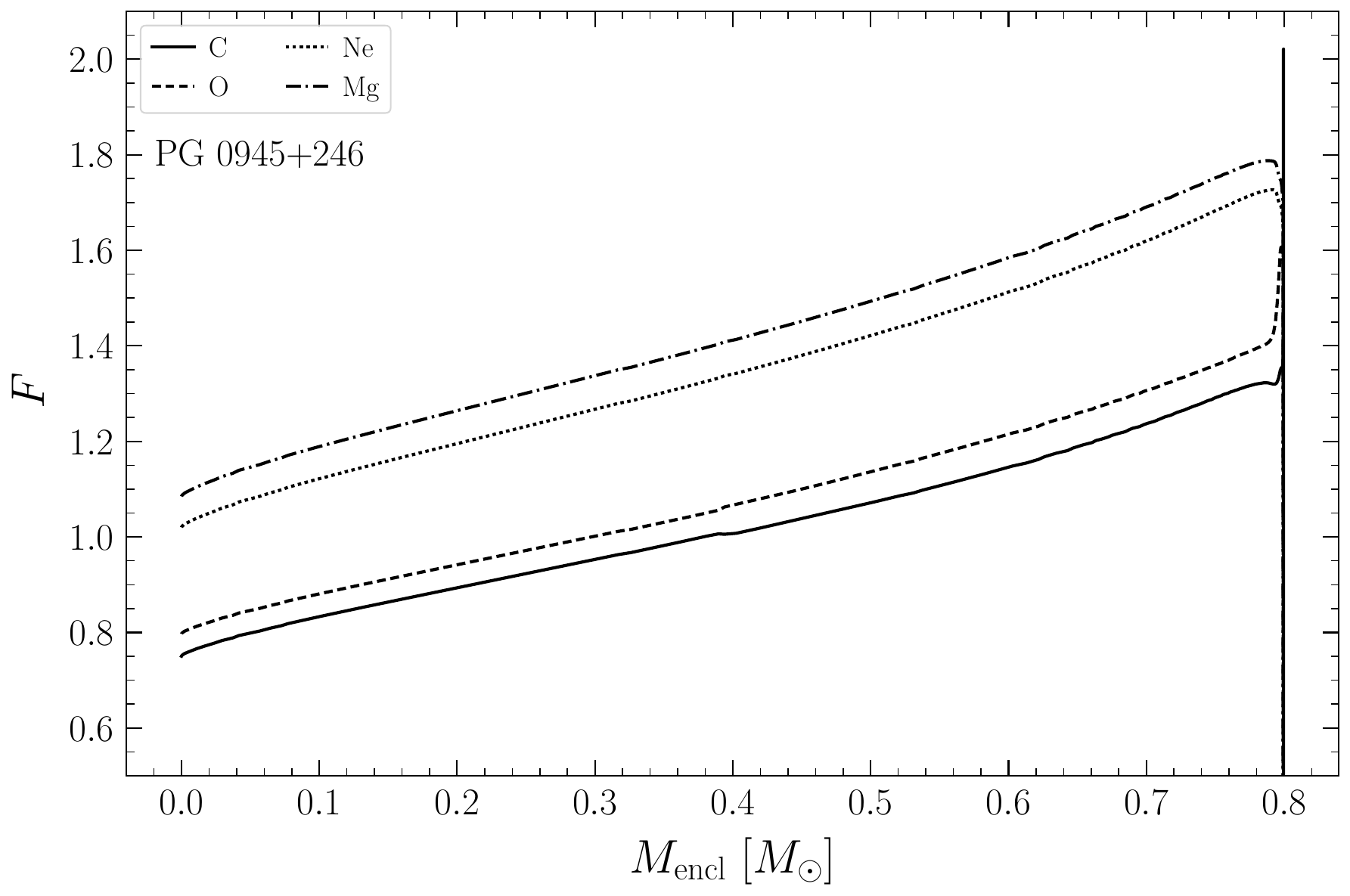}
\end{minipage}

\caption{The stellar profile of the calculated $F$-factors using our fiducial MESA model for PG 0945+246. As discussed in the main text, we adopt the parameterization used in~\cite{Nakagawa:1988rhp}.}
\label{fig:PG_FZ}
\end{figure}

We show in Fig.~\ref{fig:PG_MESA} the fiducial density and chemical abundances for our model of PG 0945+246. From this one can see that the core is predominantly carbon and oxygen, which is expected for an isolated WD of this mass. In Fig.~\ref{fig:PG_FZ} we illustrate the interior profile of the $F$-factors defined in~\eqref{eq:dedw}, which are discussed more in the main text, and in Fig.~\ref{fig:intermediate} of the main text we illustrate the summed quantity in~\eqref{eq:dedw}. Both of these quantities are relevant for the calculation of the axion emission expected from this MWD. We also note that the density of $\rho \lesssim 10^7$ g/cm$^3$ informs us that the interior of PG 0945+246 does not generally transition to the lattice phase, which otherwise would have additionally suppressed axion production. 

\begin{figure}[!htb]
\centering
\begin{minipage}{\columnwidth}
\includegraphics[width=\columnwidth]{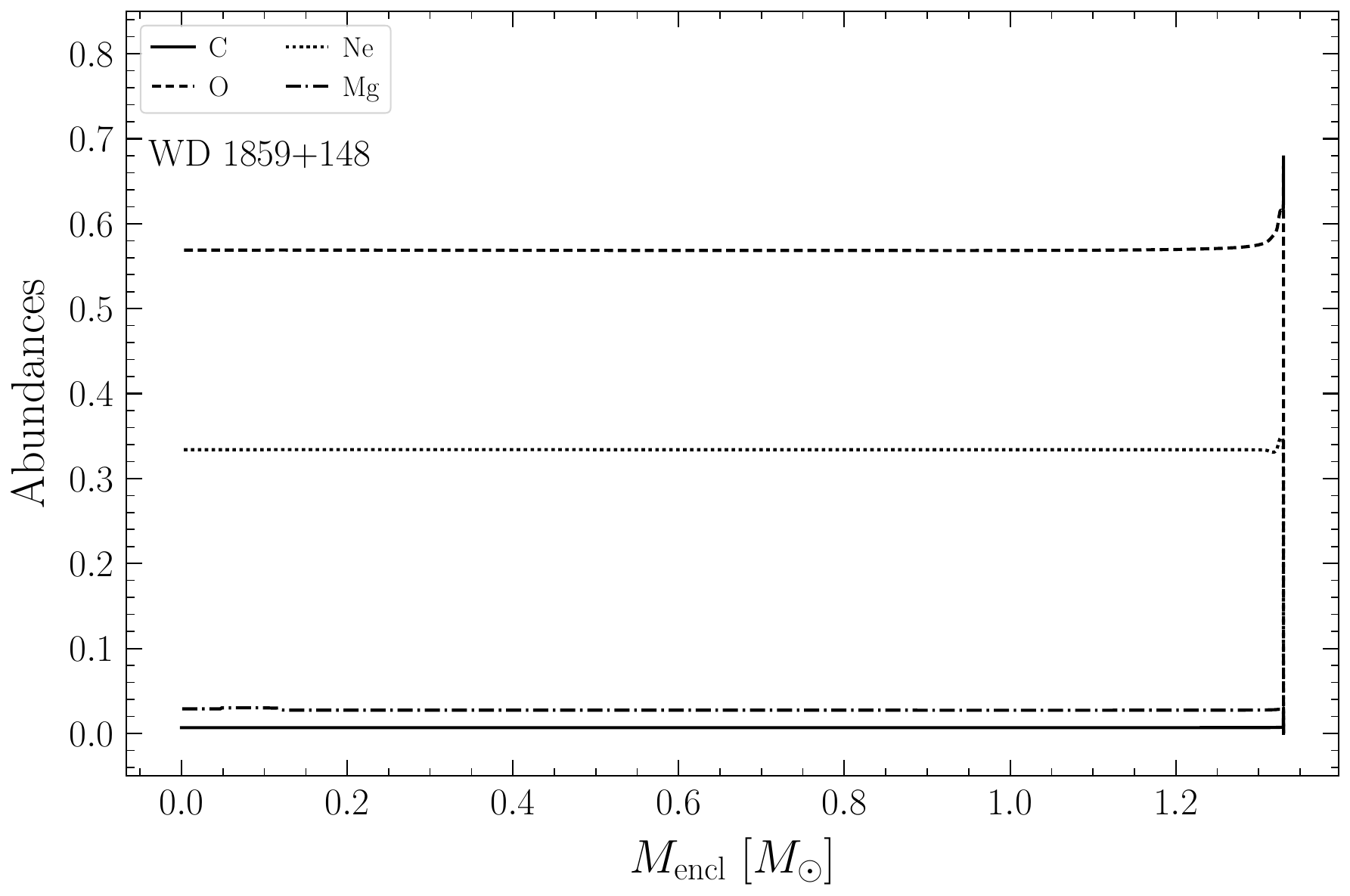}
\end{minipage}
\vspace{0.5cm}

\begin{minipage}{\columnwidth}
\includegraphics[width=\columnwidth]{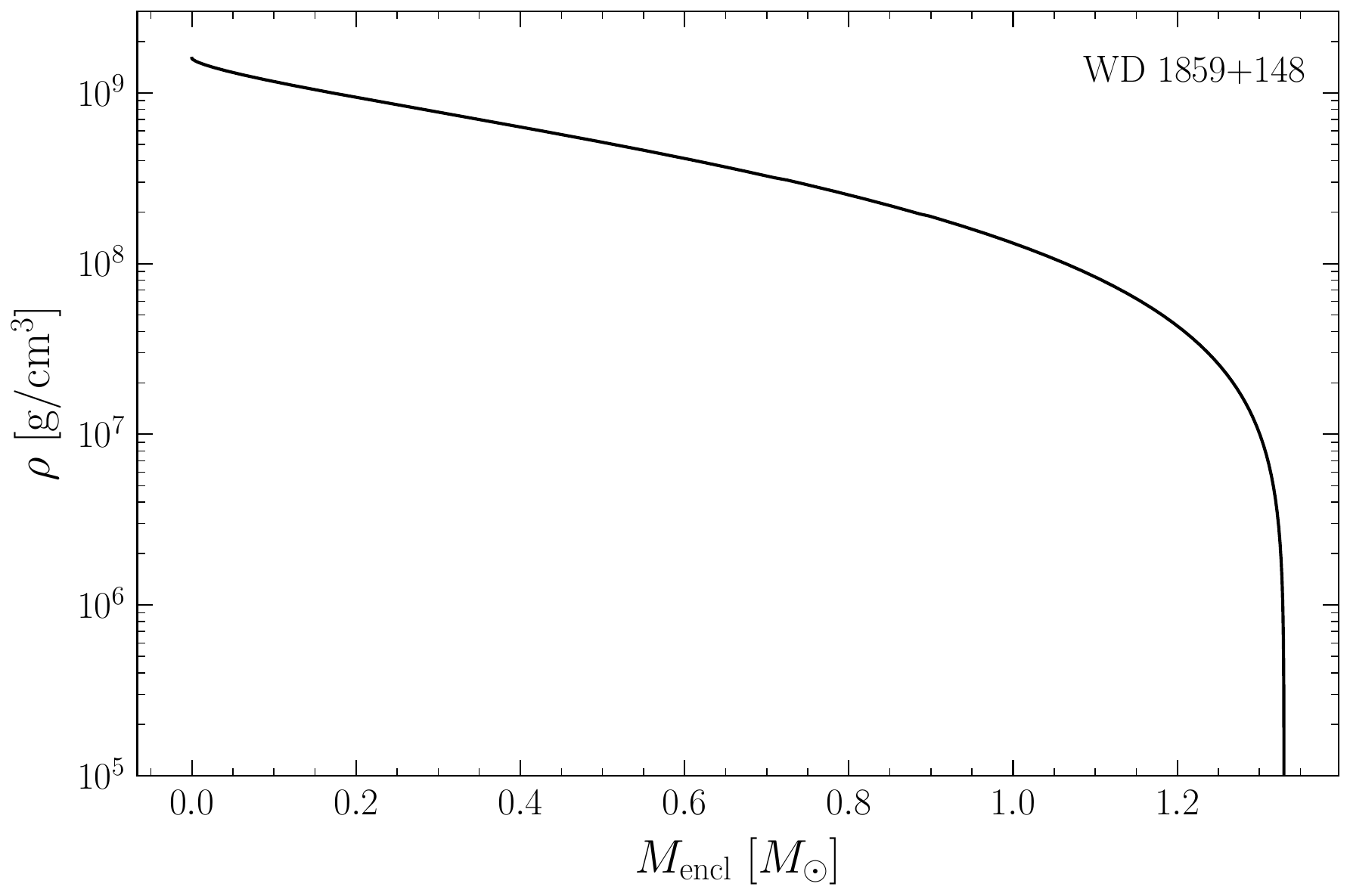}
\end{minipage}
\caption{(Top) The carbon, oxygen, neon, and magnesium chemical abundances from our fiducial MESA model for WD 1859+148, taken from the parameterized ultra-massive WD models in~\cite{2021ApJ...916..119S}, described more in the text. (Bottom) The density profile for the same model.}
\label{fig:ZTF_MESA}
\end{figure}

\begin{figure}[!htb]
\centering
\begin{minipage}{\columnwidth}
\includegraphics[width=\columnwidth]{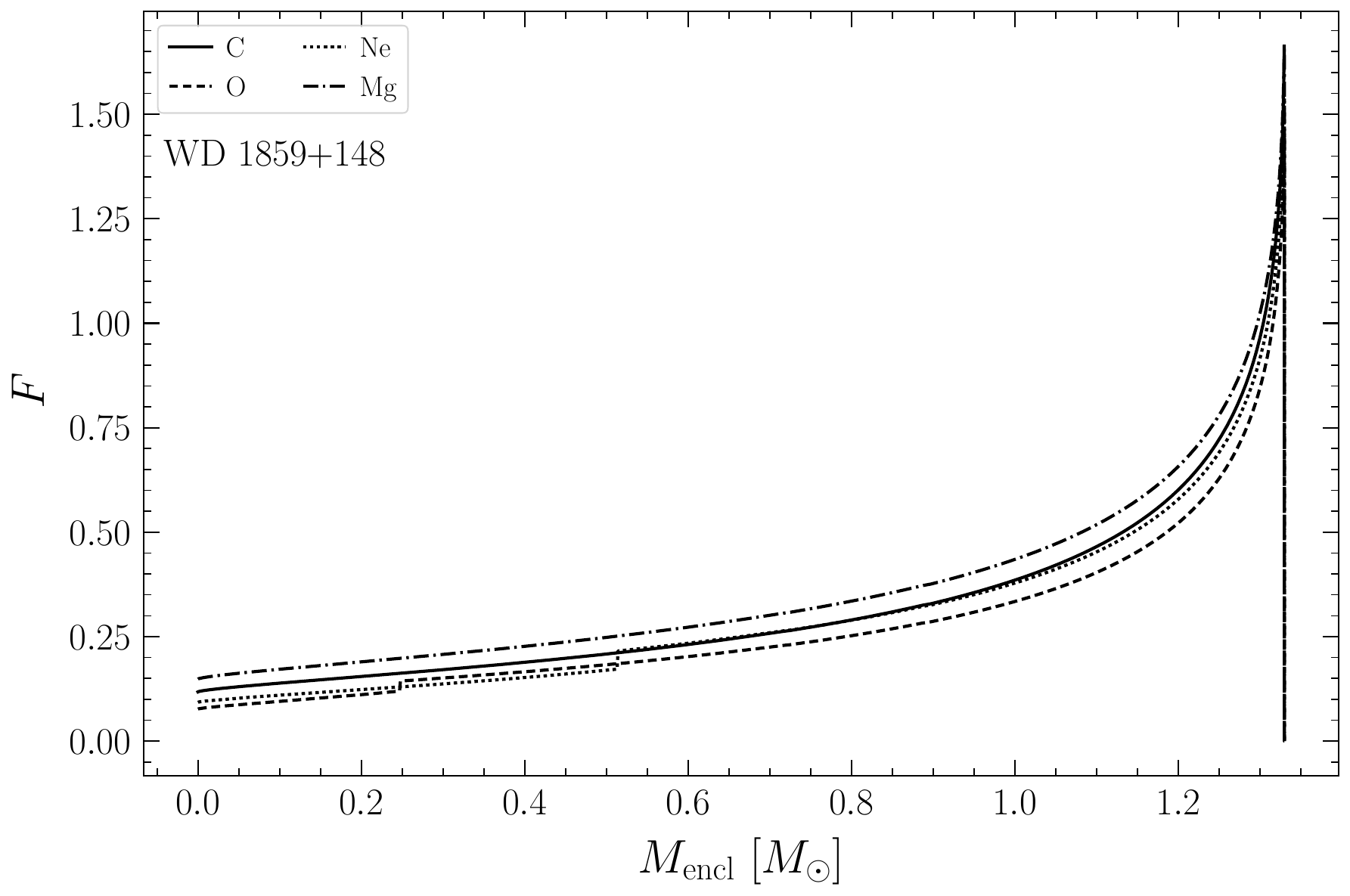}
\end{minipage}
\vspace{0.5cm}

\begin{minipage}{\columnwidth}
\includegraphics[width=\columnwidth]{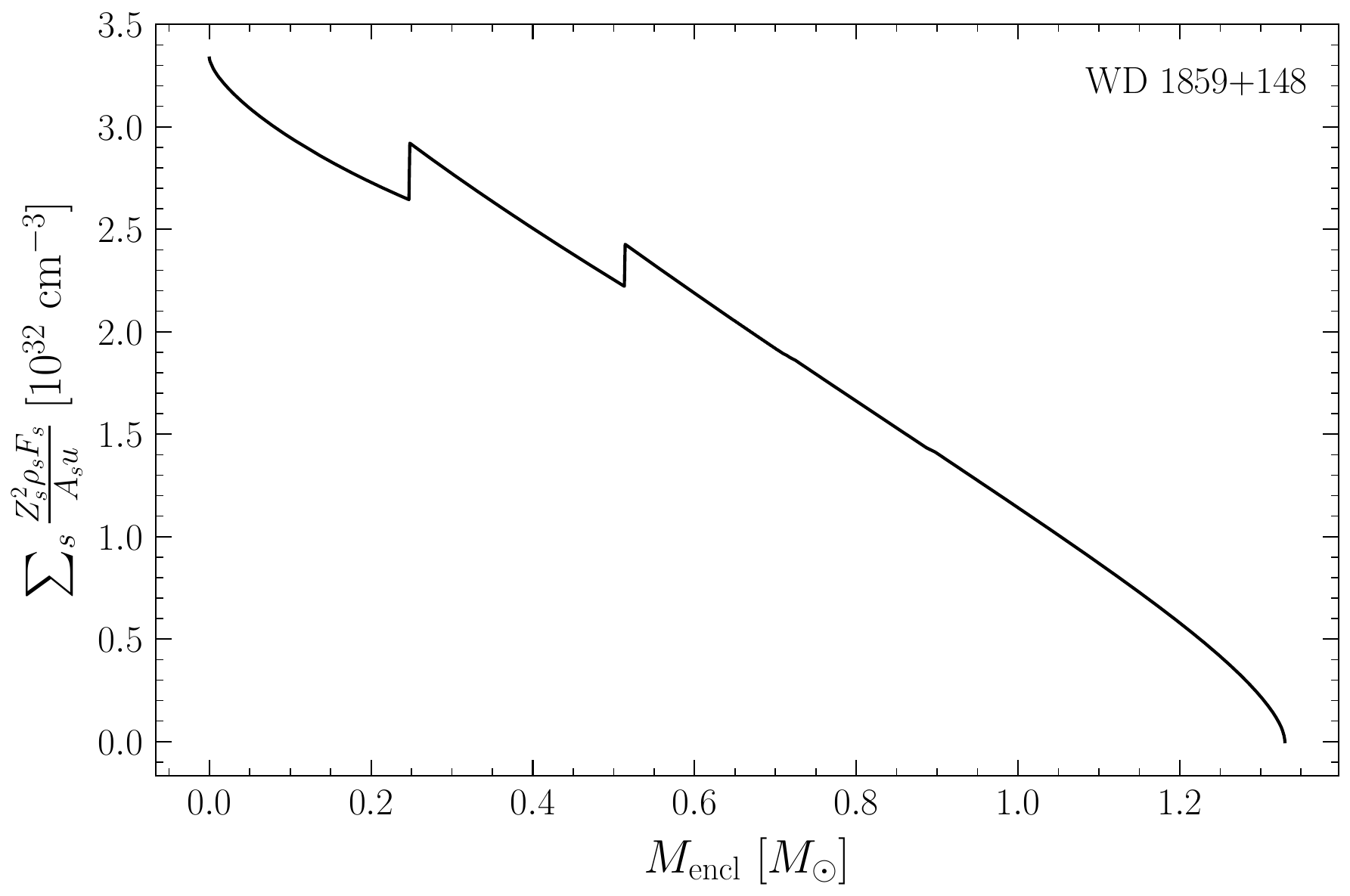}
\end{minipage}
\caption{The same as Fig.~\ref{fig:PG_FZ} but for WD 1859+148.}
\label{fig:ZTF_FZ}
\end{figure}

For WD 1859+148, a similar approach is taken in~\cite{Caiazzo:2021xkk} using multi-instrument measurements, which includes photometry from Pan-STARRS and Swift UVOT, as well as \textit{Gaia} parallax measurements. The analysis and modeling in~\cite{Caiazzo:2021xkk} is more involved, however, due to WD 1859+148's predicted high mass close to the Chandrasekhar limit; at such high masses, the core reaches very high densities, and cooling is likely to be dominated through the Urca process, which describes neutrino emission from electron capture on sodium ions~\cite{1970Ap&SS...7..374T}. This makes WD 1859+148 particularly unusual amongst other WDs previously studied~\cite{Caiazzo:2021xkk}. From the analysis in~\cite{Caiazzo:2021xkk} which takes these effects into account, we adopt the inferred estimated core temperature range $T_c = 2.2 \pm 0.4$ keV. Also discussed in~\cite{Caiazzo:2021xkk}, we utilize dedicated high-mass WD MESA simulations from~\cite{2021ApJ...916..119S}, which undertook a thorough study of the mechanisms relevant for the cooling of massive WDs ($\gtrsim 1.3$ $M_{\odot}$). Using carefully constructed sets of fixed-composition parameterized WD models which balance the difficult uncertainties in ultra-massive WD evolution with reasonable estimates~\cite{Camisassa:2022pet}, Ref.~\cite{2021ApJ...916..119S} demonstrated that cooling of these ultra-massive WDs is indeed dominated by the Urca process, and that this applies in particular to WD 1859+148. From the analysis and discussion in~\cite{2021ApJ...916..119S}, which examined WD 1859+148 in particular, we utilize their $1.33$ $M_{\odot}$ O/Ne WD MESA model as our fiducial model for WD 1859+148. We extract density and chemical abundance profiles from this model, which are illustrated in Fig.~\ref{fig:ZTF_MESA}, with intermediate quantities shown in Fig.~\ref{fig:ZTF_FZ}. 

We note that, unlike the case with PG 0945+246, WD 1859+148 reaches densities $\rho \gtrsim 10^7$ g/cm$^3$, and thus the WD interior reaches the lattice phase transition, which tends to dampen axion emissivity. The transition from the liquid phase to the lattice ion structure can be seen in the discontinuities of some of the profiles in Fig.~\ref{fig:ZTF_FZ}.

After having inferred core temperatures, stellar densities, and chemical abundance profiles, we can compute the total predicted axion luminosity from both WD 1859+148 and PG 0945+246. The spectra expected from axion bremsstrahlung, {\it i.e.}~\eqref{eq:dedw}, can be computed for each radial slice in the MESA-generated profiles. The total axion luminosity is then the integral over the star, given as the axion luminosity spectrum $dL_a/d\omega$ (in, \textit{e.g.}, erg/s/keV). Combined with the axion-photon conversion probabilities and the distances $d_{\rm WD}$ to each MWD, the total axion-induced flux $dF/d\omega$ expected at Earth for each of our MWDs is then given by
\begin{equation}
    \frac{dF}{d\omega} (\omega) = \frac{dL_a}{d\omega}(\omega) \times p_{a\to\gamma}(\omega) \times \frac{1}{4 \pi d_{\rm WD}^2} \,,
\label{eq:dFdw_total}
\end{equation}
with further details regarding this formalism found in~\cite{Dessert:2021bkv}.

\bibliography{axion}

\end{document}